\documentclass[
    reprint,
    preprintnumbers,
    superscriptaddress,
    nofootinbib,
     amsmath,amssymb,
     aps,
     prd,
    floatfix,
    ]{revtex4-2}
\usepackage{mathtools}
\usepackage{amsmath}
\usepackage{appendix}
\usepackage{comment}
\usepackage{color}
\usepackage{ulem}

\newcommand{\new}[1]{\textbf{#1}}

\begin{document}
\title{Rarity of precession and higher-order multipoles in gravitational-waves from\\ merging binary black holes}
\author{Charlie Hoy}
\affiliation{University of Portsmouth, Portsmouth, PO1 3FX, United Kingdom}
\author{Stephen Fairhurst}
\affiliation{Gravity Exploration Institute, School of Physics and Astronomy, Cardiff University, Cardiff, CF24 3AA, United Kingdom}
\author{Ilya Mandel}
\affiliation{School of Physics and Astronomy, Monash University, Clayton, VIC 3800, Australia}
\affiliation{ARC Center of Excellence for Gravitational-wave Discovery (OzGrav), Melbourne, Australia}
\date{\today}

\begin{abstract}
The latest binary black hole population estimates argue for a subpopulation of unequal component mass binaries with spins that are likely small but isotropically distributed. 
This implies a non-zero probability of detecting spin-induced orbital precession and higher order multipole moments in the observed gravitational-wave signals. In this work we directly calculate the probability for precession and higher-order multipoles in each significant gravitational-wave candidate observed by the LIGO--Virgo--KAGRA collaborations (LVK). We find that only one event shows substantial evidence for precession: GW200129\_065458, and two events show substantial evidence for higher-order multipoles: GW190412 and GW190814; any evidence for precession and higher-order multipole moments in other gravitational-wave signals is consistent with random fluctuations caused by noise. We then compare our observations with expectations from population models, and confirm that current population estimates from the LVK accurately predict the number of observed events with significant evidence for precession and higher-order multipoles.  In particular, we find that this population model predicts that a binary with significant evidence for precession will occur once in every $\sim 50$ detections, and a binary with significant evidence for higher-order multipoles will occur once in every $\sim 70$ observations. However, we emphasise that since substantial evidence for precession and higher-order multipoles have only been observed in three events, any population model that includes a subpopulation of binaries yielding $\sim 2\%$ of events with detectable precession and higher-order multipole moments will likely be consistent with the data.
\end{abstract}

\maketitle

\section{Introduction}
The LIGO--Virgo--KAGRA  collaborations (LVK) have announced a total of $90$ gravitational-wave (GW) observations~\cite{LIGOScientific:2018mvr, LIGOScientific:2020ibl, LIGOScientific:2021usb,LIGOScientific:2021djp} across three Advanced LIGO~\cite{LIGOScientific:2014pky} and Advanced Virgo~\cite{acernese2014advanced} GW observing runs; a number of additional candidates have also been reported from independent groups~\cite{venumadhav2020new, zackay2019highly, zackay2019detecting, Nitz:2021zwj,Olsen:2022pin,Wadekar:2023gea,Mehta:2023zlk,Williams:2024tna}. Considering only observations with false alarm rates (FAR) $<1\,\mathrm{yr}^{-1}$, the LVK have announced a total of 69 confident GW signals generated from binary black-hole (BBH) coalescences~\cite{LIGOScientific:2021psn}, as described in the third gravitational-wave catalogue (GWTC-3)~\cite{LIGOScientific:2021djp}.

Detailed parameter estimation routines~\cite{Veitch:2014wba,Ashton:2018jfp,Biwer:2018osg,Lange:2018pyp,Zackay:2018qdy,Smith:2019ucc,Romero-Shaw:2020owr,Ashton:2021anp,Tiwari:2023mzf} provide estimates for the binary properties given the observed GW. While the chirp mass, the mass combination that drives the phase evolution of the GW at the lowest order,~\cite{Blanchet:1995ez}, is generally well measured from observations~\cite{Tiwari:2020otp,LIGOScientific:2021psn,Tiwari:2021yvr}, the individual component masses and black hole spins remain challenging to estimate owing to strong degeneracies in the parameter space~\cite{Cutler:1994ys,Poisson:1995ef,Baird:2012cu}.

Deciphering the mass and spin distribution of binary black holes allows for constraints to be placed on their formation mechanism.  The evolution of isolated binary stars is likely to produce a population of binary black holes with spins (almost) aligned with the orbital angular momentum~\cite[e.g.][]{Kalogera:1999tq, Mandel:2009nx, Gerosa:2018wbw}, while binary black holes formed through dynamical interactions in dense stellar environments are likely to yield isotropic spin misalignment distributions~\cite[e.g.][]{Rodriguez:2016vmx,Belczynski:2014iua,Mandel:2018hfr}. Measurements of spins misaligned with the orbital angular momentum are therefore a key for deciphering between the isolated and dynamic formation channels.

A binary with spin misaligned with the orbital angular momentum undergoes spin-induced orbital precession~\cite{Apostolatos:1994mx}. Although imprints of precession have already been studied in several GW signals~\cite{LIGOScientific:2020stg,LIGOScientific:2020ufj,LIGOScientific:2021djp}, precession is generally difficult to decipher since its measurability depends on the binary's orientation, polarization, component masses and spins ~\cite{Fairhurst:2019vut,Fairhurst:2019srr,Green:2020ptm}. To date, GW200129\_065458~\cite{LIGOScientific:2021djp} (hereafter GW200129) is the only GW signal to be reported with strong evidence for spin-induced orbital precession~\cite{Hannam:2021pit}; see further discussion in Refs.~\citep{LIGOScientific:2021djp,Hoy:2022tst,Payne:2022spz,Macas:2023wiw,Kolitsidou:2024vub,Gupte:2024jfe}. However, when analysing the entire population of binary black holes observed through GWs, it has been argued that precession of the orbital plane is required to best describe the data~\cite{LIGOScientific:2021psn}.

A GW is typically decomposed into multipole moments using the $-2$ spin-weighted spherical harmonics~\citep{Goldberg:1966uu}. Binary black holes with spins aligned with the orbital angular momentum, hereafter referred to as non-precessing binaries, predominantly emit radiation in the dominant $(\ell, |m|) = (2, 2)$ quadrupole. However, additional power may also be radiated in higher order multipole moments (HoMs), including the $(\ell, |m|) = (3, 3)$ and $(4, 4)$ multipoles. Since the amplitude of each higher order term is significantly smaller than the dominant quadrupole mode, it is challenging to clearly identify additional power within the GW strain data. However, as the binary masses become more asymmetric, the relative amplitude of the HoMs increases~\cite{Mills:2020thr}; indeed many of the HoMs amplitudes vanish for equal mass binaries, particularly the $(\ell, |m|) = (3, 3)$ multipole which is otherwise the most significant for sources observed with current ground-based GW detectors~\citep{Mills:2020thr}. Since GW190412~\cite{LIGOScientific:2020stg} and GW190814~\cite{LIGOScientific:2020zkf} were the first events to have unequivocally unequal masses, they were the first signals that exhibited strong evidence for HoMs~\cite{LIGOScientific:2020stg,LIGOScientific:2020zkf,Colleoni:2020tgc,Islam:2020reh,Hoy:2021dqg}. In addition marginal evidence for HoMs was also reported in GW151226~\cite{Chia:2021mxq}, GW170729~\cite{LIGOScientific:2018mvr,Chatziioannou:2019dsz,Mateu-Lucena:2021siq}, GW190519\_153544~\cite{Hoy:2021dqg} and GW190929\_012149~\cite{Hoy:2021dqg}. Despite strong degeneracies making it challenging to infer the mass ratio of the binary, strong evidence for the presence of the $(\ell, m) = (3, 3)$ multipole was found in the population of binary black holes from the first half of the third GW observing run~\cite{Hoy:2021dqg}, while only weak preference was found for HoMs in the first gravitational-wave catalogue (GWTC-1)~\citep{Payne:2019wmy}.

Through detailed hierarchical Bayesian
analyses~\cite{Thrane:2018qnx,Mandel2019,Vitale:2020aaz}, the LVK and others have inferred that although binary black holes have a preference for comparable mass components, there is growing support for unequal component masses. It has also been reported that black hole spins are likely small, with half of the spin magnitudes below $\sim 0.25$, and spin orientations broadly isotropic with a mild preference for spins aligned with the orbital angular momentum~\cite{Tiwari:2018qch,LIGOScientific:2021psn,Tiwari:2021yvr,Vitale:2022dpa,Callister:2022qwb,Edelman:2022ydv,Callister:2023tgi}. However, other works have argued that there may be a subpopulation of black holes with lower spin magnitudes~\citep{Roulet:2021hcu,Galaudage:2021rkt,Hoy:2021rfv}, while the observed asymmetry in the effective spin distribution argues against isotropic spin orientations~\cite{Roulet:2021hcu}.

In this paper we take advantage of the harmonic decomposition of a GW~\citep{Fairhurst:2023idl} to directly calculate the probability for precession and HoMs in all significant GWs observed by the LVK. We show that only one event shows substantial evidence for precession: GW200129, and two events show substantial evidence for HoMs: GW190412 and GW190814. We therefore argue that any apparent evidence for precession and HoMs in other GW signals is consistent with random fluctuations caused by noise at the population level; for example, in a population of 100 events, we expect one GW signal to have a 1 in 100 random fluctuation from noise alone.

We then compare our observations with expectations from population models, and show that the population model from the LVK~\citep{LIGOScientific:2021psn} accurately predicts the number of observed events with significant evidence for precession and HoMs; we show that a binary with significant evidence for precession will occur once in every $\sim 50$ detections, and a binary with significant evidence for HoMs will occur once in every $\sim 70$ observations (when excluding GW190814 to remain consistent with current population estimates~\citep{LIGOScientific:2021psn}), when assuming detector sensitivities during the third gravitational-wave observing run. However, we argue that since substantial evidence for precession and HoMs has only been observed in three events to date, any population model that includes a subpopulation of binaries with $\sim 2\%$ chance of detectable precession and HoMs will likely be consistent with the data.

This paper is structured as follows: in Section~\ref{sec:method} we briefly review the process of identifying whether an observed GW has significant evidence for precession and HoMs and in Section~\ref{sec:results_pvals} we identify how many GW observations have significant evidence for precession and HoMs. We then investigate a correlation between the precision at which the binary's precession phase can be measured and its observed evidence for precession in Section~\ref{sec:phase_vs_pval}. In Section~\ref{sec:results_expectations} we then calculate the probability of detecting a GW with significant evidence for precession and HoMs based on current population estimates. We finally conclude in Section~\ref{sec:discussion} with a summary and discussion of future directions.

\section{Methods} \label{sec:method}

A quasicircular binary black hole is characterised by fifteen parameters: eight intrinsic parameters describing the component masses and spins, $[m_{1}, \mathbf{S}_{1}, m_{2}, \mathbf{S}_{2}]$, and seven extrinsic parameters describing the binary's sky location, orientation $\theta_{JN}$ (defined as the polar angle between the total angular momentum $\mathbf{J}$ and the line of sight $N$), distance, coalescence time, phase and polarization, $[\alpha, \delta, d_{L}, \theta_{JN}, $t$, \phi, \psi]$. The GW emission can be decomposed into multipoles using spin-weighted spherical harmonics.  The amount of power emitted in each multipole depends upon the properties of the system.  
A more significant fraction of power is radiated in multipole moments beyond the dominant quadrupole for binaries with unequal component masses~\citep{Mehta:2017jpq,Kalaghatgi:2019log,Mills:2020thr}. HoMs may not always be detectable in the observed GW signal since their observability depends on the binary's extrinsic parameters: HoMs are more easily detectable in binaries observed edge-on than binaries observed close to face-on~\citep{Mills:2020thr}.

As the binary black hole inspirals due to the emission of GWs, it may undergo spin-induced orbital precession leaving visible modulations in the observed GW~\citep{Apostolatos:1994mx}.
A binary black hole undergoes precession when the spin of the binary $\mathbf{S} = \mathbf{S}_{1} + \mathbf{S}_{2}$ is misaligned with the orbital angular momentum, $\mathbf{S} \times \mathbf{L} \neq 0$. In this case, both the spins and the orbital angular momentum precess around the direction of the total angular momentum. The strength of precession is typically characterised by the tilt angle of the binary's orbit, $\beta$, defined as the polar angle between $\mathbf{L}$ and the total angular momentum $\mathbf{J} = \mathbf{L} + \mathbf{S}$, or by the scalar quantity $\chi_{\mathrm{p}}$, defined in Ref.~\citep{Schmidt:2014iyl}, which parametrises the rate of precession through the spin components misaligned with the orbital angular momentum.  However, these quantities are not sufficient to unambiguously identify if the source is precessing through Bayesian methods.  For example, non-zero values of $\chi_{p}$ are often reported, even when there is no observable precession, due to the choices of priors on the spins. The visible modulations in the observed GW depend strongly on the binary's extrinsic parameters~\citep{Fairhurst:2019vut,Fairhurst:2019srr,Green:2020ptm}. 
For instance, 
precession in a binary observed close to face-on, $\theta_{JN}\sim 0$, is significantly harder to detect than precession in a binary observed edge-on, $\theta_{JN} = \pi / 2$~\citep{Apostolatos:1994mx,Brown:2012gs,Vitale:2014mka,LIGOScientific:2016ebw,Vitale:2016avz,Fairhurst:2019srr} since the leading precession contributions to the waveform vanish for face-on signals~\citep{Fairhurst:2019vut}.

\subsection{The harmonic decomposition of a GW signal} \label{sec:harmonic_decomp}

Given that the observability of precession and HoMs depends on the binary's intrinsic and extrinsic parameters, Ref.~\cite{Fairhurst:2023idl} most recently introduced the harmonic decomposition of a GW signal. This decomposition specifically isolates the leading-order contribution to the waveform from precession and HoMs and allows for their amplitudes to be calculated. Here, we briefly describe the decomposition, and refer interested readers to \cite{Khan:2019kot} and Appendix A of Ref.~\cite{Fairhurst:2023idl} for more details.

As is well known, a gravitational-wave signal can be decomposed into a series of (spin-weighted) spherical harmonics. 
For a spin-aligned system, it is natural to use a co-ordinate frame aligned with the orbital angular momentum when performing the decomposition.  For a precessing system, the emission is also most simply described in a frame aligned with the orbital angular momentum.  However, due to precession, this is not a fixed frame, but rather one that co-precesses with the binary.  It has been shown that the signal from a precessing binary is well approximated by ``twisting up''~\citep{Boyle:2011gg,Hannam:2013oca, Khan:2019kot} the multipoles of the non-precessing counterpart, based on the evolution of $\mathbf{L}$ around $\mathbf{J}$.  In effect, this twisting up procedure leads to each $(\ell, |m|)$ multipole in the co-precessing frame being split across all multipoles of the same $\ell$ in the $\mathbf{J}$-aligned frame \cite{Khan:2019kot}; for example the dominant $(\ell, |m|) = (2, 2)$ multipole in the co-precessing frame is split across all five $\ell = 2$ multipoles.  By re-expressing the waveform in terms of harmonics of specific frequency, it can be shown that each multipole can be written as sum of $2\ell + 1$ \textit{precessing harmonics}, whose frequencies differ by multiples of the precession frequency.  Individual harmonics do not exhibit precession, but the well known amplitude and phase modulations are obtained through neighbouring harmonics moving in and out of phase.  Furthermore, the amplitudes of the precessing harmonics form a power series in $b = \tan{\beta / 2}$, and for the majority of the parameter space $b \ll 1$.  In \cite{Fairhurst:2019vut}, it was demonstrated that two harmonics are sufficient to describe the $(\ell, |m|) = (2, 2)$ waveform, i.e. $h_{22}\approx h' + h_{\mathrm{p}}$.  The amplitudes of other $(\ell, |m|)$ multipoles are smaller than the $(2, 2)$ and for these the leading precession harmonic suffices \cite{Fairhurst:2023idl}.
 
As a result, a GW signal in the $\mathbf{J}$-aligned frame can be expressed as,

\begin{equation} \label{eq:harmonic_decomp}
    h \approx h' + h_{\mathrm{p}} + \sum_{(\ell, |m|)\, \neq\, (2, 2)} h_{\ell m}
\end{equation}
where $h_{\ell m}$ corresponds to the leading precessing harmonic for each HoM $(\ell, |m|) \neq (2, 2)$. Through the harmonic decomposition, we naturally see the effect of precession and HoMs through $h_{\mathrm{p}}$ and $h_{\ell m}$. This allows us to make several key observations:

\begin{enumerate}
    \item If $h_{\mathrm{p}}$ and $h_{\ell m}$ are both small in amplitude, the GW signal will be similar to the signal produced from a non-precessing binary with no observable HoMs, $h\approx h'$.
    \item If $h_{\mathrm{p}}$ is small and $h_{\ell m}$ large, the observed GW will be similar to a non-precessing binary with observable HoMs, $h\approx h' + h_{\ell m}$.
    \item If $h_{\mathrm{p}}$ is large and $h_{\ell m}$ small, the observed GW will be similar to the signal produced from a precessing binary with no observable HoMs, $h\approx h' + h_{\mathrm{p}}$. This will typically occur for a binary with close to equal component masses and high in-plane spins.
    \item If $h_{\mathrm{p}}$ and $h_{\ell m}$ are both large, the observed GW will be similar to a signal produced from a precessing binary with observable HoMs, $h\approx h' + h_{\mathrm{p}} + h_{\ell m}$.
\end{enumerate}
For this reason, it is useful to quantify the power, or signal-to-noise ratio (SNR), in $h_{\mathrm{p}}$ and $h_{\ell m}$, $\rho_{\mathrm{p}}$ and $\rho_{\ell m}$ respectively.
Intuitively, $\rho_{\mathrm{p}}$ parameterises the amount of observable precession in a GW signal, and is what we actually \textit{measure} in the data (thereby avoiding complications due to non-trivial priors). Previous studies~\cite{Green:2020ptm} have shown that $\rho_{\mathrm{p}}$ indeed correlates strongly with the measurability of precession across the parameter space, and can be used to identify when the inference on $\chi_{\mathrm{p}}$ and $\beta$ may sensitively depend on the chosen prior: when $\rho_{\mathrm{p}}$ is small, the GW signal will be similar to the signal produced from a non-precessing binary. Similarly, $\rho_{\ell m}$ quantifies the amount of observable HoM content in the signal. Given that $\rho_{\ell m}$ increases for more asymmetric binaries observed edge-on~\cite{Mills:2020thr}, when $\rho_{\ell m}$ is small, the binary must have (close-to) equal mass components or be observed approximately face-on. This information is often difficult to extract from marginalized posterior distributions alone due to parameter degeneracies, see e.g.~\cite{Baird:2012cu,Usman:2018imj}.

\subsection{Calculating the probability that the observed $\rho_{\mathrm{p}}$ and $\rho_{\ell m}$ are caused by noise}
\label{sec:calculating_probs}

We wish to calculate the probability that the SNR in precession, $\rho_{\mathrm{p}}$, and the SNR in HoMs, $\rho_{\ell m}$, for each observed GW signal is caused by noise. This allows us to identify which GW signals show significant evidence for precession and HoMs, and which are simply consistent with random fluctuations caused by Gaussian noise. We will then be able to make statements about the astrophysical implications for the individual sources, and the population as a whole.

Rather than measuring $\rho_{\mathrm{p}}$ and $\rho_{\ell m}$ directly from the GW strain data, as is done in Ref.\cite{Fairhurst:2023idl}, here we infer them from the posterior parameter estimates obtained through Bayesian inference analyses~\citep{Green:2020ptm,Mills:2020thr,Hoy:2021dqg}. This can be achieved by taking advantage of the \texttt{PESummary}~\cite{Hoy:2020vys} and \texttt{simple-pe}~\cite{Fairhurst:2023idl} libraries. In this work we use the publicly released posterior samples, made available in the GWTC-3 data release~\cite{LIGOScientific:2021djp}, and calculate $\rho_{\mathrm{p}}$ and $\rho_{\ell m}$ by using the posterior samples obtained with the \texttt{IMRPhenomXPHM}~\cite{Pratten:2020ceb} waveform model. GW200129 is the only signal where the \texttt{IMRPhenomX} waveform family is not used since it was shown that the \texttt{NRSur7dq4}~\cite{Varma:2019csw} model more accurately describes the observed GW signal~\cite{Hannam:2021pit}.

As described in Ref.\cite{Hoy:2021dqg}, the expected $\rho_{\mathrm{p}}$ and $\rho_{\ell m}$ distributions are dependent on the matched filter SNRs, $\rho^{MF}(d)$ (we hereafter drop the explicit dependence on the data $d$), of the subdominant contributions to the waveform: $h_{\mathrm{p}}$ for precession and $h_{\ell m}$ for higher order multipoles. When marginalizing the likelihood over the unknown phase, the expected posterior distribution can be shown to be,

\begin{equation} \label{eq:noise_distribution}
    p(\rho | d) \propto p(\rho)\, I_{0}(\rho^{\mathrm{MF}}\rho)\exp{\left( -\frac{[\rho^{\mathrm{MF}}]^{2} + \rho^{2}}{2} \right)} \, .
\end{equation}
where $\rho$ denotes either the precession or HoM SNR, $p(\rho)$ is the prior distribution for $\rho$, $p(\rho |d)$ is the posterior distribution given data $d$, $I_{0}$ is the Bessel function of the first kind and $\rho^{\mathrm{MF}}$ is the matched filter SNR.  The parameter estimation results provide us with the posterior distribution $p(\rho |d)$ and from this, we wish to estimate the matched filter SNR, $\rho^{MF}$.  We can then calculate the probability of obtaining $\rho^{MF}$ due to noise alone, as this follows a known distribution.

To calculate the expected SNR distribution for a given value of $\rho^{MF}$, the prior distribution $p(\rho)$ is needed. As described in Ref.~\cite{Hoy:2021dqg}, an \textit{informed prior} can be used to reduce systematic biases in the distribution. The informed prior accounts for the volume of parameter space consistent with different SNRs in $\rho_{\mathrm{p}}$ and $\rho_{\ell m}$ --- for instance observation of an equal mass binary reduces the expectation of observing HoMs and precession from noise, while a binary with unequal masses is much more likely to have observable power in precession or HoMs.
The construction of informed priors for $\rho_{p}$ and $\rho_{\ell m}$ uses results of parameter estimates that neglect precession and higher order modes respectively. 
We obtain these additional posterior samples by analysing each observed GW signal with the \texttt{IMRPhenomXHM}~\cite{Garcia-Quiros:2020qpx} and \texttt{IMRPhenomXP}~\cite{Pratten:2020ceb} waveform models. All settings were chosen to match the original LVK analyses. For the case of GW200129 we use the same settings as those described in Ref.~\cite{Hannam:2021pit}, and we obtain these additional posterior samples by analysing GW200129 with the aligned-spin limit of the \texttt{NRSur7dq4}, and \texttt{NRSur7dq4} restricted to use just the $\ell = 2$ multipoles respectively.

Once the informed prior distribution has been constructed, we infer the values of $\rho_{\mathrm{p}}^{MF}$ and $\rho_{\ell m}^{MF}$ which are consistent with the observed distributions.  In practice, we obtain the value of $\rho^{\mathrm{MF}}$ by minimizing the Jensen-Shannon divergence~\cite{61115} between the observed $p(\rho | d)$ and the inferred distribution given a specific value of $\rho^{MF}$. Finally, we assign a p-value by identifying the probability of obtaining the required value of $\rho^{\mathrm{MF}}$ from a chi distribution with 2 degrees of freedom \cite{Hoy:2021dqg}. 

Throughout this work we restrict attention to the observation of only the $(3, 3)$ HoM. This is because Refs.~\citep{Mills:2020thr,Hoy:2021dqg} most recently highlighted that only the $(3, 3)$ multipole is expected to contribute significantly to GW signals observed with current detector sensitivities. Consequently, throughout this work, we use a simplified version of Equation~\ref{eq:harmonic_decomp}: $h \approx h' + h_{\mathrm{p}} + h_{33}$. We note, however, that for binaries that are strongly precessing, or emit significant power in HoMs beyond the $(\ell, |m|) = (3, 3)$ harmonic, additional terms are needed in this harmonic decomposition~\cite{Mills:2020thr,McIsaac:2023ijd,Fairhurst:2023idl}. Although this approximation has been rigorously tested in other works~\cite[e.g.][]{Fairhurst:2023idl}, we further show the validity of the approximation for a binary with significant precession and HoM contributions in Appendix~\ref{app:assessing_harmonic_decomp}. We show that even for an extreme binary configuration, which is unlikely to be observed according to current population estimates, we see good agreement between the approximate and full GW signals.

\section{Results}

\subsection{Evidence for precession and higher order multipole moments}\label{sec:results_pvals}

\begin{figure}[t!]
    \centering
    \includegraphics[width=0.49\textwidth]{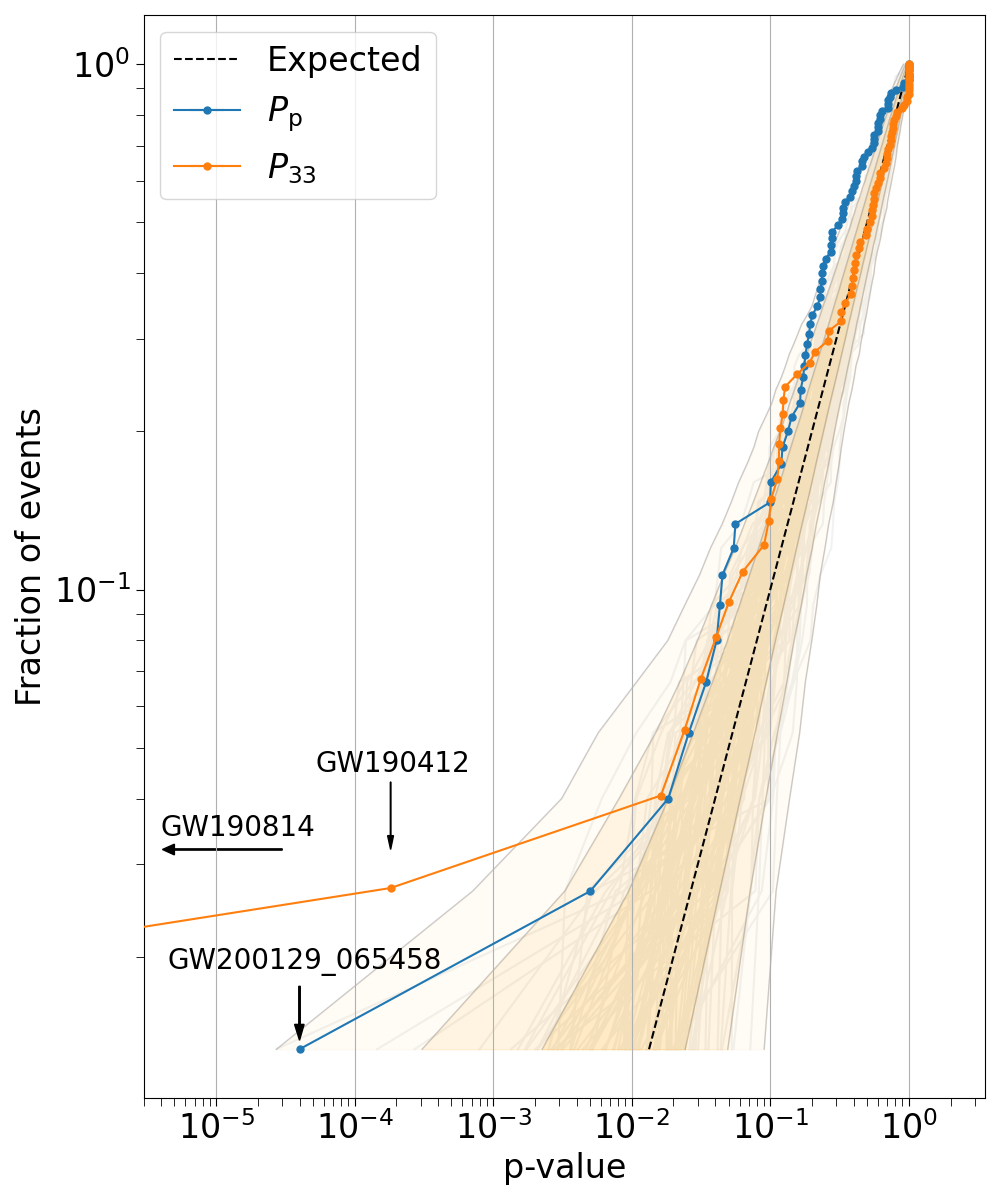}
    \caption{The cumulative distribution of p-values for the evidence of precession (solid blue) and higher order multipoles (orange) for the GW candidates reported in GWTC-3. The dashed black line shows the expected distribution of p-values under the assumption that Gaussian noise solely produces the observed precession and higher order multipole signatures in the observed GW signals, with the corresponding  $1\sigma$, $2\sigma$ and $3\sigma$ scatter shown by the orange bands. The grey lines show 100 random draws from the expected distribution under the assumption that Gaussian noise solely produces the observed precession and higher order multipole signatures in the GW signal.}
    \label{fig:pvalue}
\end{figure}

Figure~\ref{fig:pvalue} shows the cumulative distribution of p-values for precession and HoMs for all confident binary black hole mergers observed by the LVK. The dashed black line shows the expected uniform distribution of p-values under the null hypothesis that there is no signal from precession or HoMs, i.e.~under the assumption that Gaussian noise alone produced any precession and HoM signatures in the observed GWs.  We also show the expected $1\sigma$, $2\sigma$ and $3\sigma$ scatter in p-values under the null hypothesis for a population of 69 binary black holes estimated from a binomial distribution, see e.g. Ref.~\cite{LIGOScientific:2012fcp} for details). The blue and orange curves show the actual distribution of p-values for precession and HoMs, respectively.

We see that GW200129 is the only GW signal that shows strong evidence for precession. The probability that this event is non-precessing, and that Gaussian noise was responsible for the observed precession signature, is $4\times 10^{-5}$, or 1 in 25,000. This is consistent with results previously presented in Ref.~\cite{Hannam:2021pit}. For a population of 69 binary black holes this maps to an almost $3\sigma$ violation of the no-precession null hypothesis. Recently, there has been discussion concerning the impact of non-Gaussian noise artefacts impacting the precession measurement for GW200129~\cite{Payne:2022spz, Macas:2023wiw}. Since we do not account for non-Gaussianities in the GW strain data, we assume that any non-Gaussian noise artefacts in the GW strain data has been appropriately removed in our analysis. As noted in Section~\ref{sec:calculating_probs}, this result is based on posterior samples obtained with the \texttt{NRSur7dq4}~\cite{Varma:2019csw} waveform model, as described in Ref.~\citep{Hannam:2021pit}. If we use the posterior samples obtained with the \texttt{IMRPhenomXPHM}~\cite{Pratten:2020ceb}, made available through the GWTC-3 data release~\cite{LIGOScientific:2021djp}, the probability that GW200129 is non-precessing, and that Gaussian noise is responsible for the observed precession signature, increases to $9\times 10^{-4}$, or 1 in 1,100.  However, for 69 observations, the overall no-precession null-hypothesis is then violated at the low level of less than 2 $\sigma$.  Given that \texttt{IMRPhenomXPHM} infers a lower $\rho_{\mathrm{p}}$ than \texttt{NRSur7dq4}, it is not surprising that the p-value increases. However, as described in Ref.~\citep{Hannam:2021pit}, \texttt{IMRPhenomXPHM} may not meet the accuracy requirements of GW200129. We therefore only present the p-value based on samples obtained by Ref.~\citep{Hannam:2021pit}\footnote{The LVK also analysed GW200129 with the \texttt{SEOBNRv4PHM} waveform model~\citep{Ossokine:2020kjp}. If we use the posterior samples obtained with \texttt{SEOBNRv4PHM}, the probability GW200129 is non-precessing is further increased to 1 in 10. Given that an alternative sampler~\citep{Lange:2018pyp} and waveform model was used to obtain these posterior samples, and \texttt{SEOBNRv4PHM} may not meet the accuracy requirements of GW200129~\citep{Hannam:2021pit}, we do not present these results in the main text. However, we include them here for comparison.}.

GW190517\_055101 was found to be the second most significant candidate for an observation of precession, with a probability that this event is non-precessing, and that Gaussian noise is responsible for the observed precession signature, of $5\times 10^{-3}$, or 1 in 200. When compared to the expected distribution from a population of 69 non-precessing binary black holes, however, the observed precession effects are consistent with noise. GW190412 was previously  identified as a candidate with possible precession signatures~\cite{LIGOScientific:2020stg, Colleoni:2020tgc, Islam:2020reh, Mandel:2020lhv, Zevin:2020gxf, Nitz:2021uxj, Hoy:2021dqg}. Our analysis infers a probability that this event is non-precessing of 0.04, i.e. a 1 in 25 chance that the observed precession effects are caused by Gaussian noise. The precession p-values for all other GW candidates are consistent with the expected distribution from a population of 69 non-precessing binary black holes.

In Figure~\ref{fig:pvalue}, we see an excess of low-significance events that have smaller p-values for precession than expected. This trend was also seen in Ref.~\cite{Hoy:2021dqg}. Although Ref.~\cite{Hoy:2021dqg} noted some possible reasons for this, we expect that this small excess is a statistical fluctuation. In particular, we note that the Jensen-Shannon divergence is relatively flat near the minimum which means that small variations in the $\rho_p$ distribution can lead to large differences in the inferred p-value.

Only GW190814 and GW190412 show strong evidence for HoMs, specifically the $(\ell, m) = (3, 3)$ multipole, with probabilities of these events coming from sources with no HoMs of $10^{-11}$ and $2\times 10^{-4}$, respectively.  Based on the expected distribution of p-values from a population of 69 binary black holes with no HoMs, these are a $> 5\sigma$ and just over a $4\sigma$ detection, respectively. This supports the conclusions found in Refs.~\cite{LIGOScientific:2020zkf,LIGOScientific:2020stg,Islam:2020reh,Colleoni:2020tgc}. The strong evidence for HoMs indicates that both binaries have unequal component masses; indeed a full Bayesian analysis infers mass ratios $\sim 9:1$ and $\sim  3.5:1$ respectively~\cite{LIGOScientific:2020zkf,LIGOScientific:2020stg}. The strong evidence for HoMs in GW190412 has also allowed for measurements of the sources recoil velocity~\citep{CalderonBustillo:2022ldv}.

\begin{figure}[t!]
    \centering
    \includegraphics[width=0.49\textwidth]{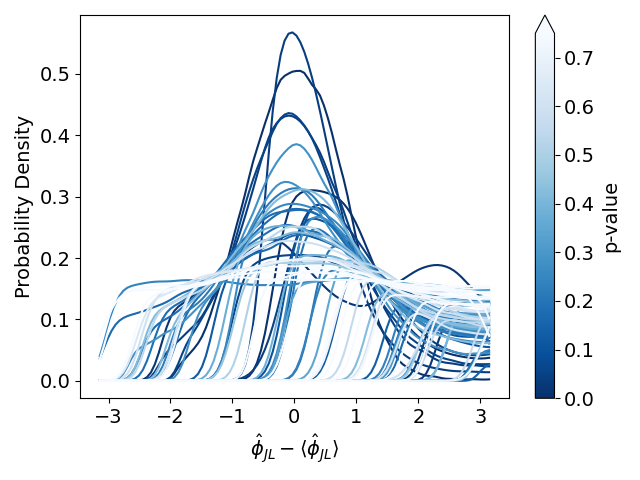}
    \caption{Posterior distributions for the precession phase, $\hat{\phi}_{JL}$~\cite{Roulet:2022kot}, for the observed binary black hole mergers in GWTC-3, coloured by the p-value for the evidence of precession. As in Ref.~\cite{Roulet:2022kot}, we artificially centre each posterior distribution by subtracting the circular mean.}
    \label{fig:phi_jl}
\end{figure}

GW190929\_012149 is the next most significant candidate for an observation of HoMs with a probability of no HoM content in the underlying source of $0.02$, i.e. 1 in 50 chance that Gaussian noise is responsible for the observed HoM effects. Although this was deemed marginal in Ref.~\cite{Hoy:2021dqg}, the larger population presented here means that this is now in line with expectations. Marginal evidence for HoMs was previously reported in GW151226~\cite{Chia:2021mxq}, GW170729~\cite{LIGOScientific:2018mvr,Chatziioannou:2019dsz,Mateu-Lucena:2021siq} and GW190519\_153544~\cite{Hoy:2021dqg}. Our analysis finds that these events have p-values $0.3$, $0.1$ and $0.05$ respectively. When comparing to a population of 69 events, we see no statistical evidence for HoMs in GW151226, GW170729 and GW190519\_153544.

\subsection{Correlation between the precession phase and probability of precession} \label{sec:phase_vs_pval}

The precession phase%
\footnote{The precession phase, $\phi_{JL}$, is sometimes denoted in the literature as $\alpha$. We denote the $2\pi$-periodic version of the precession phase by $\hat{\phi}_{JL}$ to match the notation used in Ref.~\cite{Roulet:2022kot}.}
, $\phi_{JL}$, tracks the azimuthal angle of the orbital angular momentum $\mathbf{L}$ around the total angular momentum $\mathbf{J}$, and is typically defined at a reference frequency $f = 20\, \mathrm{Hz}$%
\footnote{In this work we use the same settings as the original LVK analyses. This means that for most of the confident binary black hole mergers considered, we use a reference frequency of $f = 20\, \mathrm{Hz}$. However, 22 candidates used a lower reference frequency: $f \in [3, 5, 10, 16]\, \mathrm{Hz}$. For details, see the GWTC-3 data release~\cite{LIGOScientific:2021djp}.}
When precession is not observable, the precession phase cannot be constrained and Bayesian analyses will return an uninformative measurement of $\phi_{JL}$. However, if the GW has observable precession and no HoMs, Equation~\ref{eq:harmonic_decomp} reduces to $h\approx h'  + h_{p}$, and the precession phase can be measured independently from the orbital phase. Since the orbital phase remains the same for both waveform components, while the contribution from the precession phase differs, constraints can be placed on the value of $\phi_{JL}$.

Recently Ref.~\cite{Roulet:2022kot} found that the precession phase, or rather a $2\pi$-periodic version of the precession phase $\hat{\phi}_{JL}$\footnote{Although the precession phase is typically defined at a certain reference frequency, the $2\pi$-periodic version of the precession phase, introduced in Ref.~\cite{Roulet:2022kot}, is independent of the frequency specified since it considers the difference between the phase and the circular mean of the phase.}, is recovered surprisingly well for several candidates, most notably GW200129, GW190412 and GW151226. In Figure~\ref{fig:phi_jl}, we reproduce the analysis of Ref.~\cite{Roulet:2022kot} but look for correlations between the precession p-value and the precision of the measurement of $\hat{\phi}_{JL}$.

As expected we see that, in general, a smaller precession p-value correlates with a tighter constraint on $\hat{\phi}_{JL}$. GW200129, GW190412 and GW151226 have the tightest constraints on $\hat{\phi}_{JL}$, and they have small precession p-values: $4\times 10^{-5}$, $0.04$ and $0.12$. We see one noticeable exception,  GW191109\_010717. This candidate has a p-value of 0.018 with a relatively poorly constrained $\hat{\phi}_{JL}$. 

\subsection{Comparing observations with expectations from population estimates} \label{sec:results_expectations}

\begin{figure*}[t!]
    \centering
    \includegraphics[width=0.94\textwidth]{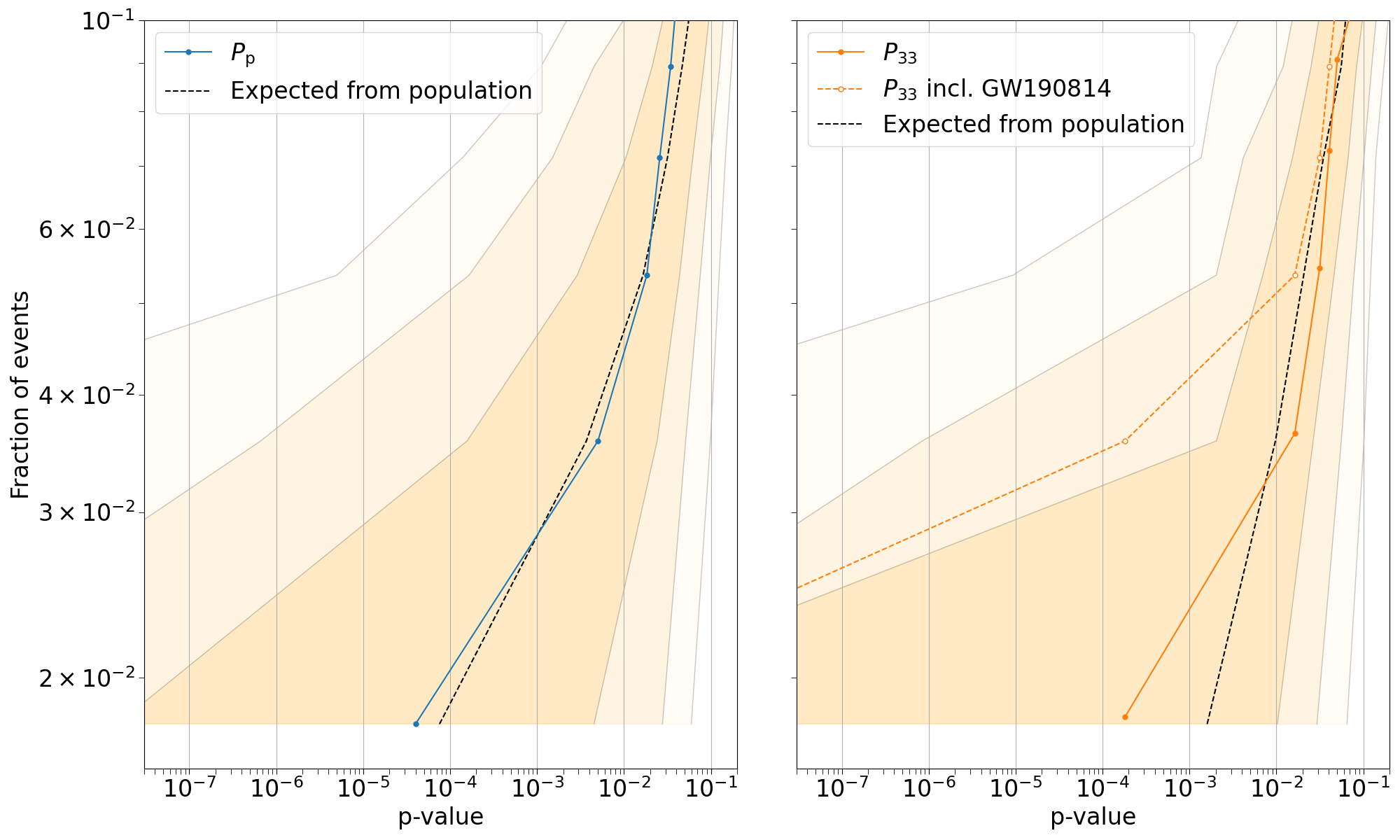}
    \caption{Cumulative distribution of p-values for the evidence of precession (left) and higher order multipoles (right) for the GW candidates observed in GWTC-3. The dashed black line shows the expected distribution of p-values as the median of random draws of 69 binaries from a mock data set sampled from the LVK population model~\citep{LIGOScientific:2021psn}, and the orange bands shown the $1\sigma$, $2\sigma$ and $3\sigma$ scatter. Since current population estimates exclude GW190814~\citep{LIGOScientific:2021psn}, we similarly remove GW190814 from the cumulative distribution shown by the solid blue and orange lines. However, we also show the cumulative distribution of p-values for the evidence of higher order multipoles when including GW190814 by the dashed orange line for comparison.}
    \label{fig:pvalue_population}
\end{figure*}

We now turn to comparing the fraction of observed events that display signatures of precession or HoMs with expectations from the population model described below.  We check for consistency by sampling binaries from the population model, calculating the expected distributions of p-values for precession and HoMs, and comparing these distributions with observations. In this work we assume a {\sc{Powerlaw+peak}} mass model~\citep{LIGOScientific:2018jsj}, a {\sc{Default}} spin model~\citep{Talbot:2017yur} and a {\sc{Powerlaw}} redshift model~\citep{Fishbach:2018edt}, all described in Appendix B of Ref.~\citep{LIGOScientific:2021psn}. We randomly draw 1000 samples from the posterior distribution on the hierarchical model parameters obtained in Ref.~\citep{LIGOScientific:2021psn}; in effect we simulate 1000 different \textit{Universes}, each consistent with the data. We then randomly sample merging binaries from each Universe. We hereafter refer to this population model as the LVK population model.

\begin{figure*}[t!]
    \centering
    \includegraphics[width=0.485\textwidth]{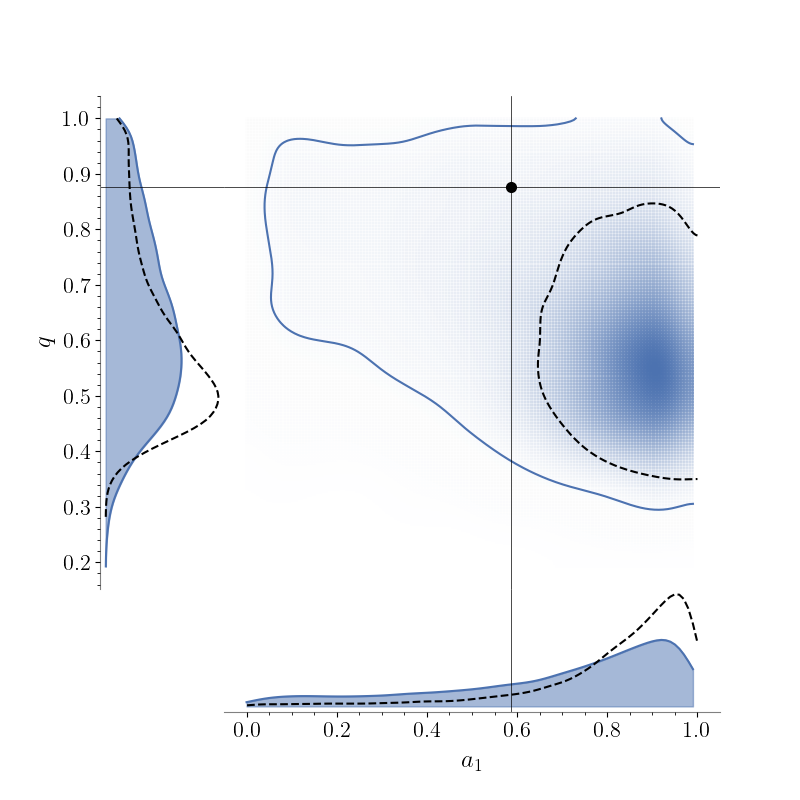}
    \includegraphics[width=0.485\textwidth]{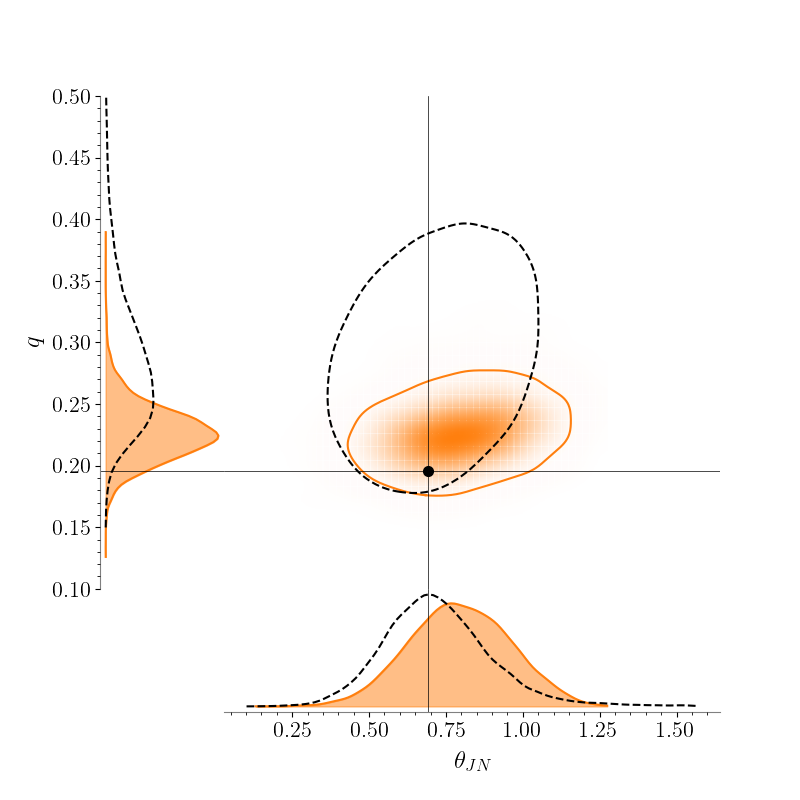}
    \caption{\textit{Left}: Posterior distribution for the inferred primary spin magnitude, $a_{1}$, and mass ratio, $q = m_{2} / m_{1} \leq 1$, for a binary chosen from the population model which was inferred to have significant evidence for precession. In black we show the inferred posterior distribution for GW200129\_065458~\citep{Hannam:2021pit} \textit{Right}: Posterior distribution for the inferred inclination angle (folded to $[0, \pi/2]$), $\theta_{JN}$, and mass ratio, $q$, for a binary chosen from the population model which was inferred to have significant evidence for higher order multipole moments. In black we show the inferred posterior distribution for GW190412~\citep{LIGOScientific:2021djp}. In both plots, the contours show the 90\% credible interval, and the cross hairs show the true values of the binaries chosen from the population model.}
    \label{fig:highly_significant_posterior}
\end{figure*}

Not all binaries drawn from population models are observable with current template-based search pipelines adopted by the LVK~\citep{Cannon:2011vi,
Privitera:2013xza,Messick:2016aqy,Hanna:2019ezx,
Sachdev:2019vvd,Usman:2015kfa,
Nitz:2017svb,Nitz:2018rgo,pycbc-software,2016CQGra..33q5012A,SPIIR2,
2012CQGra..29w5018L,2018CoPhC.231...62G}. 
Current template-based search pipelines adopted by the LVK use templates which restrict spins to be aligned with the orbital angular momentum and neglect the presence of HoMs (although see Refs.\cite{McIsaac:2023ijd,Schmidt:2024jbp,Harry:2017weg,Chandra:2022ixv,Wadekar:2023gea} for possible solutions for template-based search pipelines, Refs.~\cite{Klimenko:2008fu, Lynch:2015yin, Skliris:2020qax} for search pipelines that do not use templates, and Ref.~\cite{Chandra:2020ccy} which demonstrates that non-template based search pipelines may be more sensitive to precessing binaries than current template-based search pipelines). This means that current template-based searches are not equally sensitive to all binaries, and binaries with significant evidence for precession and HoMs may remain undetected.

In this work we only retain binaries from population models that are observable with template-based search pipelines at current detector sensitivities. We achieve this by calculating the detection probability as a function of the binary's mass and spin, and only retaining draws with high detection probabilities. We estimate the detection probability by adding a suite of randomly chosen simulated signals to real GW data and assessing the performance of template-based search pipelines across the parameter space. Given the computational cost of performing this analysis, we use the publicly available results produced by the LVK~\cite{LIGOScientific:2021psn,ligo_scientific_collaboration_and_virgo_2021_5546676}.

As discussed in Section~\ref{sec:method}, multiple Bayesian inference analyses are needed to calculate p-values for precession and HoMs for a single randomly chosen binary. Since performing Bayesian inference is a computationally expensive task, let alone analysing each binary with multiple models, we only analyse randomly chosen binaries with $\rho_{\mathrm{p}} > 1$ and $\rho_{33} > 1$. Although we expect similarities between populations with $\rho_{\mathrm{p}} > 1$ and $\rho_{33} > 1$, for example asymmetric component masses~\citep{Green:2020ptm,Mills:2020thr}, we treat them as separate populations in this work. This means that we separately draw two populations: one where binaries have $\rho_{\mathrm{p}} > 1$ and another where binaries have $\rho_{33} > 1$. We find that  $11.5\%$ and $5.6\%$ of the detectable binaries sampled from the LVK population model have $\rho_{\mathrm{p}} > 1$ and $\rho_{33} > 1$, respectively. To ensure a consistent number of binaries in each population, we initially draw $100$ detectable binaries from each Universe in the LVK population model, giving a total of $10^{5}$ randomly chosen binaries, and downsample to 1000 and 2000 binaries for the precession and HoM analyses respectively. We therefore estimate the expected distribution of p-values for precession and HoMs from population models based on $\sim 120$ simulated GW signals. We take care to renormalize the p-value distributions, based on the expected number of binaries with $\rho_{\mathrm{p}} > 1$ and $\rho_{33} > 1$.

We simulate the GW signals for each randomly chosen binary with the \texttt{IMRPhenomXPHM} waveform model~\citep{Pratten:2020ceb}, and inject the signals into expected Gaussian noise based on a representative power spectral density (PSD) from the third LVK GW observing run (O3). We then perform our Bayesian analyses with the \texttt{Dynesty} nested sampler~\citep{Speagle:2019ivv} (via \texttt{bilby}~\citep{Ashton:2018jfp,Romero-Shaw:2020owr}) with 1000 live points, and we use sufficiently wide and uninformative priors (see Appendix B.1 in Ref.~\cite{LIGOScientific:2018mvr} for details). All analyses integrated the likelihood from $20\mathrm{Hz}$ to $972.8\mathrm{Hz}$, and we used the \texttt{bilby}-implemented rwalk sampling algorithm with an average of 60 accepted steps per MCMC for all analyses.

Figure~\ref{fig:pvalue_population} compares the cumulative distribution of p-values for precession and HoMs with random draws from population estimates. The current LVK population model excludes GW190814~\citep{LIGOScientific:2021psn}, so we also excluded this event from the cumulative distribution. We see that the LVK population model accurately predicts the number of observed events with significant evidence for precession and HoMs, with the distribution of observed p-values lying comfortably within the $1\sigma$ scatter from the population samples. We see large error bars associated with the expectation from population estimates, with the most significant binaries having a large spread of p-values: $10^{-8} < P_{\mathrm{p}} < 4\times 10^{-3}$ for precession and $10^{-11} < P_{33} < 10^{-2}$ for the $(\ell, |m|)=(3, 3)$ multipole, with ranges quoted at the $1\sigma$ level. This is a consequence of analysing a limited number of randomly chosen binaries owing to computational cost. Unsurprisingly, when we include GW190814 in the cumulative distribution, the observed distribution now lies outside of the $1\sigma$ uncertainty. This is expected since current population estimates have vanishingly small probabilities for drawing binaries with extreme mass ratio configurations.

In Figure~\ref{fig:highly_significant_posterior} we show the posterior distribution for the inferred primary spin magnitude, $a_{1}$, and mass ratio, $q = m_{2} / m_{1} \leq 1$, for a binary chosen from the population model. This binary was chosen \new{among several possible candidates ($21$ out of the $\sim 120$ simulated GW signals considered)} as it was inferred to have a significant precession p-value: $P_{\mathrm{p}} < 4\times 10^{-5}$, and hence a more significant p-value than GW200129. We see that the inferred posterior distribution resembles GW200129, with near extreme spin and $q\sim 0.5$; GW200129 was inferred to have $a_{1} = 0.9^{+0.1}_{-0.5}$ and $q = 0.6^{+0.4}_{-0.2}$~\citep{Hannam:2021pit}. Figure~\ref{fig:highly_significant_posterior} also shows the posterior distribution for another randomly chosen binary from the population model; this binary was \new{again chosen among several possible candidates} as it was inferred to have a significant HoM p-value: $P_{\mathrm{33}} < 2\times 10^{-4}$, and hence a more significant p-value than GW190412. Here, we see that the inferred inclination angle, $\theta_{JN}$, and mass ratio, $q$, resembles GW190412; GW190412 was inferred to have $\theta_{JN} = 0.71^{+0.31}_{-0.24}$ and $q = 0.28^{+0.12}_{-0.07}$~\citep{LIGOScientific:2020stg}. In both cases, the true values lie within the inferred 90\% credible intervals, but there is a small bias, which is likely a consequence of the specific Gaussian noise realisation chosen. Figure~\ref{fig:highly_significant_posterior} illustrates that GW200129-like and GW190412-like signals are possible in the LVK population model, highlighting that e.g. the observed precession and HoM effects previously reported in GW200129 are consistent with a gravitational-wave signal produced from a quasi-circular merger in the presence of Gaussian noise -- non-Gaussian noise artefacts may not be needed to explain the observed posterior distributions~\cite{Payne:2022spz,Gupte:2024jfe}. 

Finally, we can use the distribution of p-values obtained from random draws from the LVK population model to calculate the probability of observing binaries with significant evidence for precession or HoMs. We find that there is a 1 in 50 chance that a detectable binary from the LVK population model will have evidence for precession that is at least as significant as for GW200129, and a 1 in 70 chance that a detectable binary will have evidence of HoMs that is at least as significant as for GW190412. Both are consistent with the observation of 1 binary in 68 (removing GW190814).

In this work, we chose to only compare our observations with one population model. However, we note that since we have only observed a single event with significant evidence for precession, and two events with significant evidence for HoMs, any population model that includes a subpopulation of binaries with a $\sim 2\%$ chance of detectable precession and HoMs will likely be consistent with our observations. This includes the LVK population model, but this is not the only possible model consistent with the data, and care should be taken in making strong statements about the population-wide distribution of spin-orbital misalignments and mass ratios based on a handful of events.

\section{Discussion} \label{sec:discussion}

In this work we investigated whether binary black holes observed by the LVK show evidence for precession and HoMs. We took advantage of the harmonic decomposition of a GW~\citep{Fairhurst:2023idl} and directly calculated the probability that the observed SNR in precession and HoMs can be explained by Gaussian noise alone. Although the SNRs in precession and HoMs have been used to help explain the physics of several GW candidates~\cite{LIGOScientific:2020stg,LIGOScientific:2020zkf}, they are generally underutilised in population analyses.

We found that only one event shows substantial evidence for precession: GW200129.  Meanwhile, two events show substantial evidence for HoMs: GW190412 and GW190814. This implies that any evidence for precession and HoMs in other GW signals is consistent with random fluctuations caused by noise. We then compared our observations with expectations from population models to ensure consistency. We found that the current LVK population model~\citep{LIGOScientific:2021psn} accurately predicts the number of observed events with significant evidence for precession and HoMs. We found that a binary with significant evidence for precession will occur once in every $\sim 50$ detections, and a binary with significant evidence for HoMs will occur once in every $\sim 70$ observations if sources are drawn from the LVK population model. Both fractions are consistent with observations, and the latter is consistent with forecasts from Monte Carlo simulations~\citep{Roy:2019phx}. However, we note that since significant evidence for precession and HoMs has only been confidently observed in three events, any population model that includes a subpopulation of binaries with $\sim 2\%$ chance of detectable precession and HoMs will likely be consistent with the data. 

Although we only find one candidate with significant evidence for precession, this does not necessarily mean that most of the observed binary black holes are non-precessing. Rather it means that the majority of the observed binary black holes \textit{do not have individually-observable precession}. This could indicate a true lack of precession in binaries whose components spins are aligned with the orbital angular momentum; or component spins that are too small in magnitude for precession to be observable despite misalignment; or extrinsic parameters that are unfavourable for the detectability of precession.
Consequently, all we can conclusively say about individual events is that we have observed precession in one candidate, GW200129. GW200129 is therefore likely formed through dynamic encounters (although there are cases where the isolated channel can produce binaries with high spins and significant precession~\citep{Steinle:2020xej}). A possibility is that GW200129 was formed through a hierarchical merger, where spin-induced orbital precession is likely; one component is predicted to be highly spinning and misaligned with the orbit. Indeed, signatures of hierarchical mergers may have already been found among GW observations~\citep{Tiwari:2020otp,Kimball:2020qyd,Antonini:2024het}. Other options include triple black hole systems~\citep{Antonini:2017tgo}, or formation in active galactic nuclei~\citep{Tagawa:2020dxe}. Given that we cannot make any statements about precession of the other individual GW candidates, we are unable to comment on their formation mechanism. We note that recently Ref.~\cite{Schmidt:2024hac} conducted a dedicated search for precessing binaries using data from O3, and did not find any evidence for additional GW candidates, despite an increase in sensitivity to precessing systems. This similarly implies that the majority of the observed binary black holes do not have individually-observable precession. We also highlight that Ref.~\citep{CalderonBustillo:2024akj} equivalently found that precession is only individually identified in GW200129 (using another observable).%

We only find two candidates with significant evidence for HoMs, both of which have conclusively unequal component masses. Given that GW190412 does not have observable precession, the isolated formation channel is possible: although the isolated formation channel favours comparable component masses~\citep{Sigurdsson:1993zrm,Rodriguez:2016kxx}, binaries with $q \leq 0.5$ are only disfavoured by a factor of $\sim 10$ and therefore unsurprising given the current population~\citep{Dominik:2012kk,Dominik:2014yma,Giacobbo:2017qhh}. Hierarchical mergers are also possible, although we would then have expected to observe significant evidence for precession. Both isolated binary evolution models~\citep{Dominik:2012kk,Dominik:2014yma,Mandel:2020} and dynamical formation scenarios~\citep{Ye:2019xvf} struggle to reproduce the extreme mass ratio of GW190814. All binaries radiate power in HoMs%
\footnote{Some HoMs (for example all odd $m$ multipoles) are suppressed for non-spinning, exactly equal mass ratio binaries. However, this is not the case for e.g. precessing binaries.}
, so the lack of observable power in most of the GW candidates is partly a consequence of their extrinsic parameters.

Throughout this work we only considered the observability of precession and HoMs in binary black holes observed with ground-based detectors operating at sensitivities. Previous works have investigated their observability for next generation detectors, such as Einstein Telescope~\citep{Maggiore:2019uih} and Cosmic Explorer~\citep{Reitze:2019iox}, and found that HoMs in particular will be observable in a larger fraction of detected GW signals~\citep{Divyajyoti:2021uty,Fairhurst:2023beb,rosenberg2024assessing}.

We predict that Einstein Telescope will observe HoMs once in every 20 candidates, while Cosmic Explorer will observe HoMs once in every 10 candidates. 
Einstein Telescope will observe precession once in 5 candidates, while Cosmic Explorer will observe precession once in every 4 candidates. To obtain these estimates, we calculate how often randomly chosen binaries from the current population models have $\rho_{\mathrm{p}}$ and $\rho_{\ell, m}$ greater than 2.1 for different detector configurations, following the analysis in  Ref.~\citep{Fairhurst:2019srr}.  While it is appropriate to use the inferred mass and spin populations from current observations \cite{LIGOScientific:2021psn}, the current observations are not capable of probing the redshift evolution of the merger rate.  Thus we draw binaries according to a representative redshift distribution for next generation detectors, such as the distributions proposed in Refs.~\citep{Ng:2020qpk,vanSon:2021zpk}, and use a low frequency cut-off of 5Hz for the SNR calculation. We also assume that all candidates with SNR greater than 8 are detectable\footnote{Although this will likely not be a significant issue for binary black holes, this calculation is not taking into consideration the motion of the GW detector during the observed signal due to the Earth's rotation.}. We do not assume a network of next generation detectors, unlike Ref.~\citep{Divyajyoti:2021uty}, so these predictions are a conservative lower bound. We see that Cosmic Explorer is more likely to observe HoMs and precession in GW signals than Einstein Telescope.  Current population estimates \cite{LIGOScientific:2021psn} indicate that the majority of black hole binaries are low mass, with $m_{1} \approx 10 M_{\odot}$.  From Figs~11 and 12 in Ref.~\cite{Fairhurst:2023beb} we see that both Cosmic Explorer and Einstein Telescope are sensitive to these mergers out to $z \gtrsim 20$, but that Cosmic Explorer has greater sensitivity to HoMs.  This is consistent with our estimate that CE will observe HoMs in a greater fraction of events than ET.

The methods presented in this paper straightforwardly and clearly identifies which GW candidates show significant evidence for precession and HoMs. This method can also be used to verify whether population estimates are consistent with observations.  We therefore suggest that future population analyses make use of this approach.

\section{Acknowledgements}

\begin{figure*}
    \centering
    \includegraphics[width=0.87\textwidth]{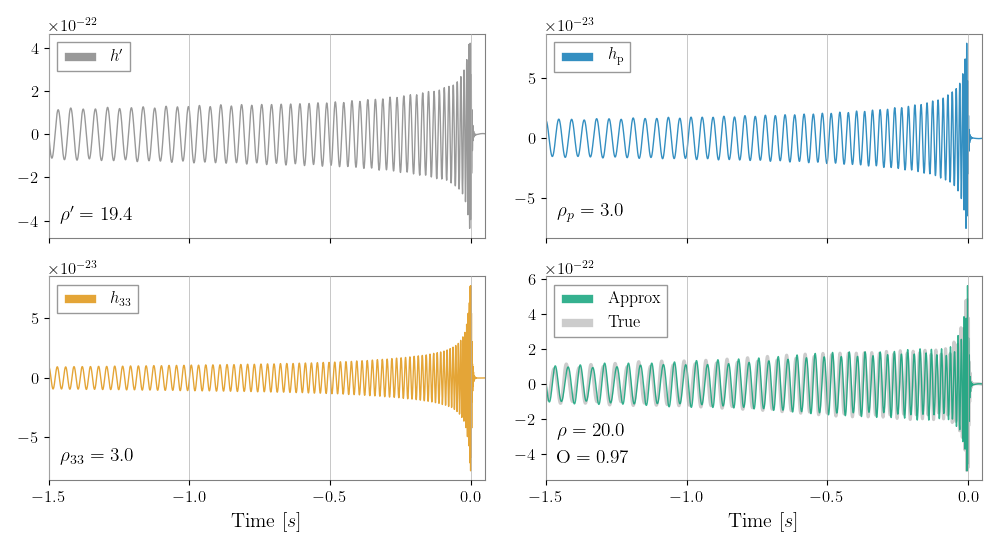}
    \caption{
    A GW signal with a signal-to-noise ratio of 20 emitted from the merger of two black holes with masses $40\, M_{\odot}$ and $10\, M_{\odot}$. The top left, top right and bottom left panels show $h', h_{\mathrm{p}}$ and $h_{33}$ respectively. The bottom right panel compares the approximate gravitational wave signal constructed by adding $h', h_{\mathrm{p}}$ and $h_{33}$, with the true signal generated with the \texttt{IMRPhenomXPHM} model~\cite{Pratten:2020ceb}. Assuming a typical PSD from the third GW observing run, the SNR in $h_{\mathrm{p}}$ and $h_{\mathrm{33}}$ is $\rho_{\mathrm{p}} = 3.0$ and $\rho_{33} = 3.0$ respectively and the overlap between the true waveform and the approximate waveform during the inspiral is $\mathrm{O} = 0.97$.
    }
    \label{fig:decomp}
\end{figure*}

We are grateful to Laura Nuttall for continued discussions throughout this project, and to Eleanor Hamilton for valuable comments on this manuscript. We thank Sylvia Biscoveanu, Juan Calder\'on Bustillo, Thomas Callister, Mark Hannam, Cameron Mills, Alex Nitz, Soumen Roy, Stefano Schmidt, Eric Thrane, John Veitch and Barak Zackay for discussions and suggestions. CH thanks the UKRI Future Leaders Fellowship for support through the grant MR/T01881X/1. SF thanks the Science and Technology Facilities Council (STFC) for support through the grant ST/V005618/1. IM acknowledges support from the Australian Research Council (ARC) Centre of Excellence for Gravitational Wave Discovery (OzGrav), project number CE230100016.  We are grateful for computational resources provided by Cardiff University, LIGO Laboratory and the ICG, SEPNet and the University of Portsmouth, and supported by STFC grant ST/N000064 and National Science Foundation Grants PHY-0757058 and PHY-0823459. This research made use of data, software and/or web tools obtained from the Gravitational Wave Open Science Center (\href{https://www.gw-openscience.org}{https://www.gw-openscience.org}), a service of LIGO Laboratory, the LIGO Scientific Collaboration and the Virgo Collaboration. LIGO is funded by the U.S. National Science Foundation. Virgo is funded by the French Centre National de Recherche Scientifique (CNRS), the Italian Istituto Nazionale della Fisica Nucleare (INFN) and the Dutch Nikhef, with contributions by Polish and Hungarian institutes. This material is based upon work supported by NSF's LIGO Laboratory which is a major facility fully funded by the National Science Foundation.

Plots were prepared with Matplotlib~\cite{2007CSE.....9...90H} and {\texttt{PESummary}}~\cite{Hoy:2020vys}. Parameter estimation was performed by {\texttt{LALInference}}~\cite{Veitch:2014wba}, {\texttt{bilby}}~\cite{Ashton:2018jfp,Romero-Shaw:2020owr} and {\texttt{pBilby}}~\cite{Smith:2019ucc} (which made use of the {\texttt{dynesty}} nesting sampling package~\cite{Speagle:2019ivv}). {\texttt{NumPy}}~\cite{harris2020array}, \texttt{pycbc}~\cite{pycbc-software} and {\texttt{Scipy}}~\cite{2020SciPy-NMeth} were also used in our analysis.

\appendix

\section{Assessing the harmonic decomposition of GWs} \label{app:assessing_harmonic_decomp}

As described in Section~\ref{sec:method}, the harmonic decomposition of GWs allows us to explicitly isolate the precessing and HoM components of observed GWs. For most of the parameter space, a GW signal can be well approximated as a sum of the leading three waveform components, $h \approx h' + h_{\mathrm{p}} + h_{33}$ where $h_{\mathrm{p}}$ is the precessing component and $h_{33}$ corresponds to the $(\ell, |m|) = (3, 3)$ HoM contribution. 

In Fig~\ref{fig:decomp} we compare the true and approximate waveforms for a binary with masses $40\, M_{\odot}$ and $10\, M_{\odot}$, and explicitly show $h', h_{\mathrm{p}}$ and $h_{33}$.  This binary is precessing with component spin magnitudes of $\chi=0.5$, both tilt angles between the spin and the orbital angular momentum equal to $\theta=0.8$ radians, and is viewed at an inclination angle of $0.5$ radians. When generating the true waveform we use the \texttt{IMRPhenomXPHM}~\cite{Pratten:2020ceb} phenomenological model, which includes the effects of both precession and HoMs and is regularly used by the LVK. We generate $h_{\mathrm{p}}$ and $h_{33}$ using code available in the \texttt{PESummary}~\cite{Hoy:2020vys} and \texttt{simple-pe}~\cite{Fairhurst:2023idl} libraries.

From Fig~\ref{fig:decomp} we see that $h'$ is the dominant contribution to the waveform with SNR $19.4$. $h_{\mathrm{p}}$ and $h_{33}$ have smaller amplitudes, with SNRs $\rho_{\mathrm{p}} = 3.0$ and $\rho_{33} = 3.0$. The total SNR of the approximate waveform $h'+h_{\mathrm{p}}+h_{33}$ is $19.8$, which is slightly smaller than the true waveform's SNR of 20. This difference is attributed to additional power in HoMs beyond $(\ell, |m|) = (3, 3)$. We find that although the GW is written with only 3 components, the sum provides a good approximation to the true GW. This implies that Eq.~\ref{eq:harmonic_decomp} is valid even for regions of the parameter space that are unlikely assuming the current population of binary black holes~\cite{LIGOScientific:2021psn}. We quantify this agreement with the overlap between the true waveform, $g$, and the approximate waveform, $h$, defined as,

\begin{equation}
    \mathrm{O} = \frac{(h | g)}{|h||g|},
\end{equation}
where,

\begin{equation}
    (h | g) = 4 \, \mathrm{Re}\int{df \frac{\tilde{h}(f)\tilde{g}^{*}(f)}{S(f)}},
\end{equation}
$\tilde{h}(f)$ represents the Fourier transform of $h(t)$, $*$ implies a complex conjugate, and $S(f)$ is the power spectral density. An overlap\footnote{Note the overlap does not maximise over phase and time.} close to unity implies that the waveforms are very similar, while a lower overlap implies that there are significant differences between the waveforms. Assuming a typical PSD from O3, we find that the overlap between $g$ and $h$ is $0.97$ when considering only the inspiral (the harmonic decomposition is more accurate during the inspiral than during the merger and ringdown). This overlap reduces to 0.95 when including the merger and ringdown.

Current population models show a preference for comparable mass components with half of the spin magnitudes below $~0.25$~\cite{Tiwari:2018qch,LIGOScientific:2021psn,Tiwari:2021yvr,Vitale:2022dpa,Callister:2022qwb,Edelman:2022ydv,Callister:2023tgi}. Therefore, the impact of precession and HoMs will be significantly reduced compared with the system shown in Fig~\ref{fig:decomp}. We therefore expect the overlap to be closer to 1.0 for the observed population of binary black holes. Indeed, when randomly drawing 1,000 observable binaries from population models, assuming current detector sensitivities, we find an average overlap between the approximate and full waveforms to be $0.98$. We therefore expect Eq.~\ref{eq:harmonic_decomp} to be a good approximation to the true GW for the majority of binary black holes observed by the LVK.

\bibliography{references}

\end{document}